\documentclass[prl,aps,final,twocolumn,superscriptaddress,floatfix,showpacs]{revtex4-1}
\usepackage{times}
\usepackage{pstricks}
\usepackage{graphicx}
\usepackage{gensymb}
\begin{document}

\title{Size dependent phase diagrams of Nickel-Carbon nanoparticles}

\author{Y. Magnin}
\affiliation{Centre Interdisciplinaire de Nanoscience de Marseille, Aix-Marseille University and CNRS, Campus de Luminy, Case 913, F-13288, Marseille, France.}

\author{A. Zappelli}
\affiliation{Centre Interdisciplinaire de Nanoscience de Marseille, Aix-Marseille University and CNRS, Campus de Luminy, Case 913, F-13288, Marseille, France.}

\author{H. Amara}
\affiliation{Laboratoire d'Etude des Microstructures, ONERA-CNRS, BP 72, F-92322, Ch\^{a}tillon, France.}

\author{F. Ducastelle}
\affiliation{Laboratoire d'Etude des Microstructures, ONERA-CNRS, BP 72, F-92322, Ch\^{a}tillon, France.}

\author{C. Bichara}
\affiliation{Centre Interdisciplinaire de Nanoscience de Marseille, Aix-Marseille University and CNRS, Campus de Luminy, Case 913, F-13288, Marseille, France.}

\begin{abstract}
The carbon rich phase diagrams of nickel-carbon nanoparticles, relevant to catalysis and catalytic chemical vapor deposition synthesis of carbon nanotubes, are calculated for system sizes up to about 3 nanometers (807 Ni atoms). A tight binding model for interatomic interactions drives the Grand Canonical Monte Carlo simulations used to locate solid, core/shell and liquid stability domains, as a function of size, temperature and carbon chemical potential or concentration. Melting is favored by carbon incorporation from the nanoparticle surface, resulting in a strong relative lowering of the eutectic temperature and a phase diagram topology different from the bulk one. This should lead to a better understanding of the nanotube growth mechanisms.
\end{abstract}

\maketitle

\begin{figure}[htb!]
\includegraphics[width=8cm]{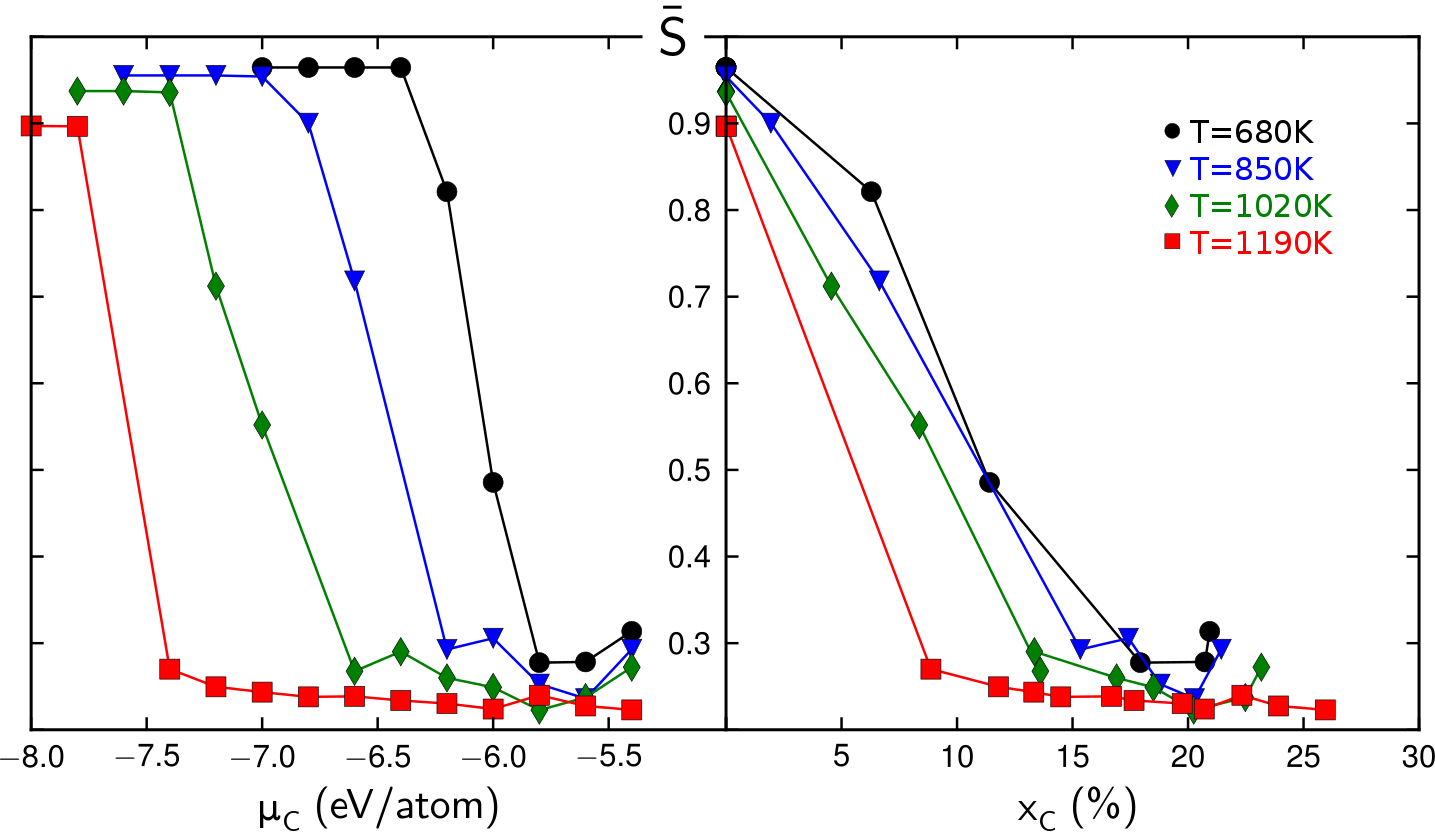}
  \caption{(Color online) Average values of the order parameter ($\bar{S}$), plotted as a function of $\mu_C$ (left) and $x_C$ (right) for a NP, initially icosahedral with 309 Ni atoms, at different temperatures. On the left panel, the solid ($\bar{S} \geq 0.85$) and liquid domains ($\bar{S}\leq 0.35$), separated by a gradual transition zone, are easily identified. The corresponding atomic compositions are readily obtained on the right panel, and a phase diagram can be constructed using the procedure described in  \cite{SI}.}
  \label{fig:fig1}
\end{figure}

\begin{figure}[htb!]
\includegraphics[width=8.5 cm]{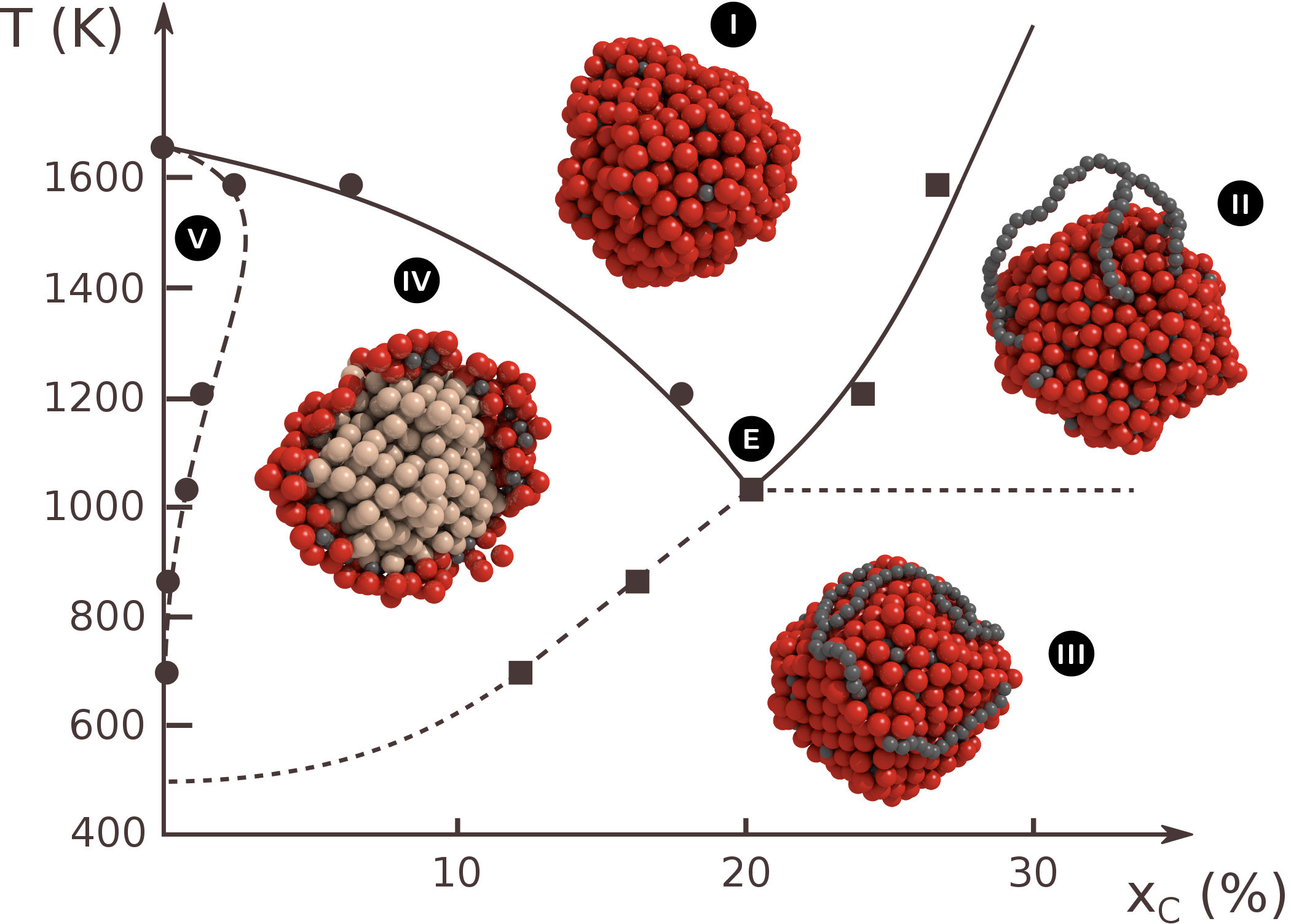}
  \caption{(Color online) Ni-C phase diagram for a NP with 807 Ni atoms. Five domains are identified, illustrated by a snapshot of a characteristic configuration (C atoms: grey, other colors: Ni atoms). I: homogeneous liquid solution; II:  C segregation at the surface of liquid NP; III: C segregation from mostly solid NP; IV: solid core / liquid shell NP; V: solid solution.}
  \label{fig:fig2}
\end{figure}

\begin{figure}[htb!]
\includegraphics[width=9cm]{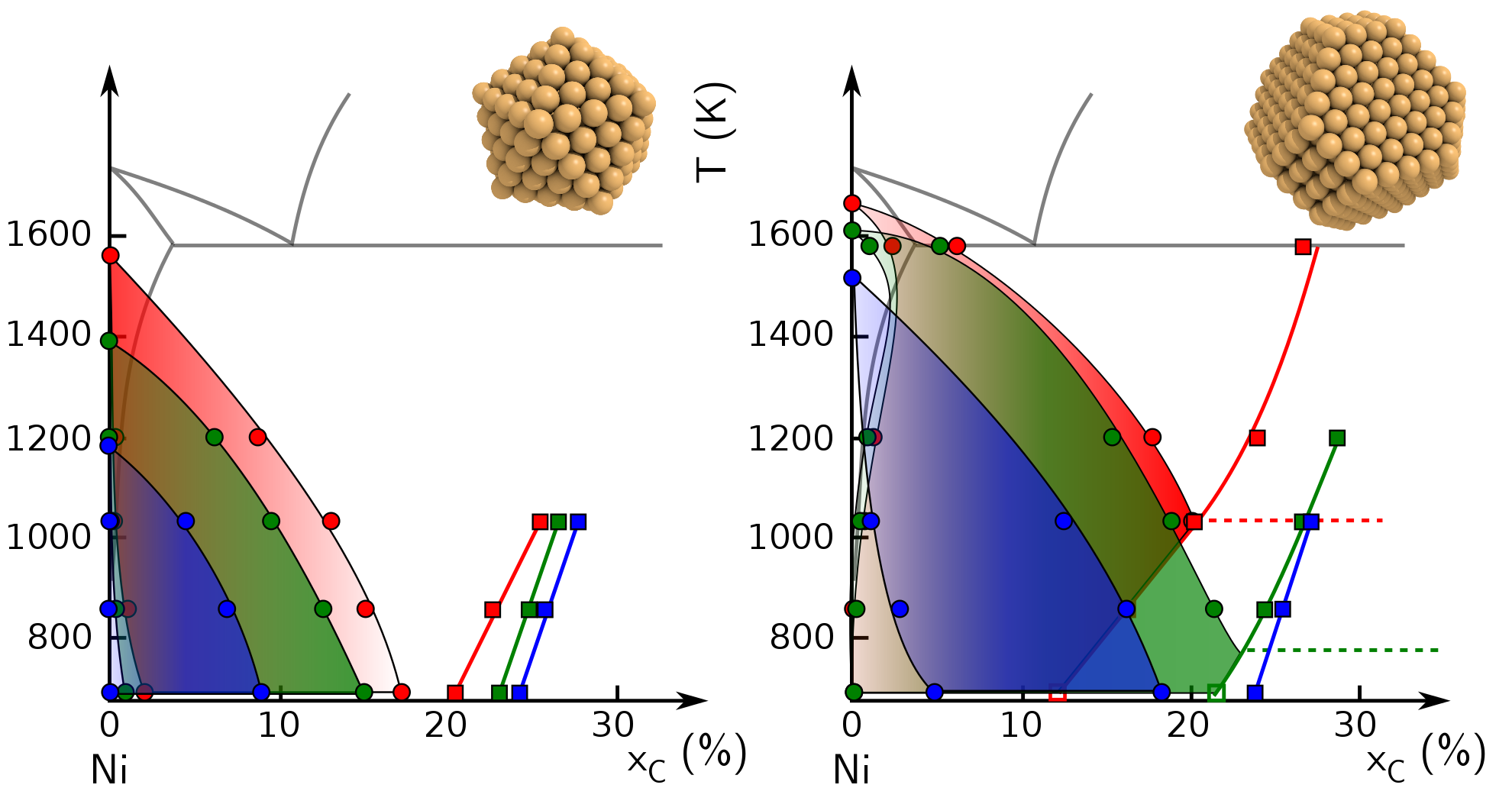}
\caption{(Color online) Size dependent Ni-C phase diagrams for icosahedral (left) or face centered cubic (Wulff shaped, right) NPs. Full black lines show the bulk phase diagram \cite{portnoi10}.\\
Left panel: NPs with 55 (blue), 147 (green) and 309 (red) atoms. Right panel: NPs with 201 (blue), 405 (green) and 807 (red) atoms. For systems smaller than Ni$_{405}$, the eutectic point, if it exists, is below $680 K$.}
  \label{fig:fig3}
\end{figure}

\begin{figure}[htb!]
\includegraphics[width=8cm]{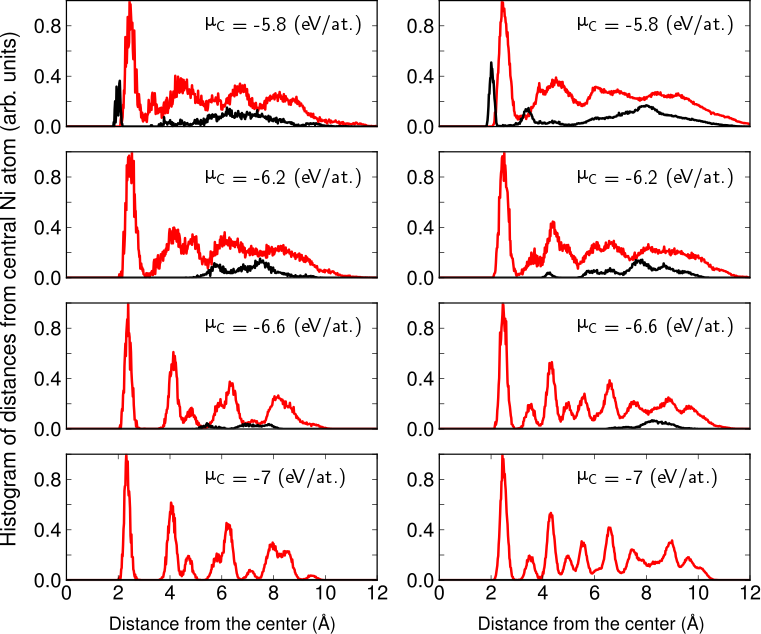}
  \caption{(Color online) Unweighted pair distribution functions (Ni-Ni, red and Ni-C, black) centered on the Ni atom closest to the center of the NP, for a Ni$_{309}$ (left) icosahedral NP and Ni$_{405}$ (right) $fcc$ Wulff-shaped NP calculated at 850 K. The Ni-C distributions have been multiplied by 2 for legibility.  Differences in crystalline arrangements of pure NPs are obvious: for $\mu_C$ = -7.0 eV/atom: the first peak lies at 0.231 nm for the icosahedral NP while it is 0.013 nm further out for the $fcc$ one, at 0.244 nm, as a result of the larger internal pressure in the former. We clearly see that an outer Ni layer melts for $\mu_C$ between -6.6 and -6.2 eV/atom, due to interstitial C incorporation, before full melting at higher $\mu_C$, when C has diffused throughout the NP.}
  \label{fig:fig4}
\end{figure}

  Reducing dimensions of materials to the nanoscale has a deep impact on their structure and properties. Back in 1976, Buffat and Borel \cite{buffat76} measured a large decrease of the melting temperature of Au nanoparticles (NP) for sizes down to 2.0 nm. Since then, stimulated by the interest for the synthesis of nanowires and carbon nanotubes by Catalytic Chemical Vapor Deposition (CCVD), phase diagrams of alloyed NPs have been actively investigated. Although the smallest catalyst NPs of typical catalyst such as Fe, Co or Ni are probably liquid under growth conditions \cite{jourdain13}, the possibility of a synthesis from solid NPs  remains open for larger ones and for other less common elements or alloys. Illustrating this dilemma, a route to chiral selective growth of Single Wall Carbon Nanotubes (SWNT) has recently been proposed \cite{yang14}, relying on  reportedly solid state NPs, in this instance CoW nanoalloys.\\
Different from elements of the same group like Si and Ge that form substitutional alloys with transition metals typically used as catalysts, carbon is smaller and forms interstitial alloys. \textit{In situ} studies are scarce, but subsurface carbon incorporation appears as an important step of the catalytic process for Pd \cite{teschner08} or Ni \cite{rinaldi11,weatherup14}. For Fe, various carbide phases are observed \cite{mazzucco14}, displaying contrasted catalytic activities. It is therefore important to understand how carbon incorporation in the catalyst NP, together with its size reduction, modifies its physical and chemical states as compared to the bulk alloy. To address this question, we  proceed to establish size dependent nickel-carbon NPs phase diagrams in a size range ($\sim$ 1-3 nm) relevant for SWNT growth.\\
  Although not in thermodynamic equilibrium,  NPs often have a long enough life time to make an experimental  determination of their phase diagram possible, usually by  Transmission Electron Microscopy (TEM) and structural investigations. Pt-Ru \cite{hills03}, Au-Ge \cite{sutter08} and Cu-Ni \cite{park08} NPs have been studied. In the latter, a phase diagram was established using  TEM experiments and CALPHAD type calculations, including interfacial Gibbs energy contributions that become important at the nanoscale. This approach is reported to remain valid for NP diameters down to about 10 nm \cite{sim14,sopousek14,guisbiers11}. In the above cited results, the topology of the phase diagrams is preserved even for very small NPs. In such a case, though, Gibb's phase rule \cite{hillert} might be no longer valid, and so does the presence of a eutectic three phase equilibrium. \\
  An alternative approach to solve the issues raised by very small NP sizes, is the direct computer simulation of their structure. Density Functional Theory (DFT) based Molecular Dynamics (MD) calculations were used for NPs up to 641 atoms \cite{rollmann07} to study the size dependent structural changes of icosahedral ($I_h$) Fe clusters, as well as FePt and CoPt nanoalloys \cite{gruner08}. However, studying binary alloy NPs involves sampling the chemical order. This requires a Monte Carlo (MC) simulation approach, associated with simple enough atomic interaction models \cite{baletto05}. Lattice models have been used to study order-disorder phase diagrams of substitutional alloys \cite{pohl12,alloyeau09,delfour09}, but interstitial alloys such as the metal-carbon systems of interest here require considering all degrees of freedom, including atomic relaxations induced by carbon incorporation in interstitial sites. In the framework of carbon nanotube growth,  the melting of Fe-C  \cite{ding04} and Ni-C clusters \cite{engelmann14} has been studied, but not the states below the liquidus lines. Assuming an equivalence between the NP size reduction and an external pressure increase on the corresponding bulk alloy, Harutyunyan \textit{et al.} \cite{harutyunyan08}, predicted a reduced carbon solubility in Fe-C NPs.\\
  In this Letter, we build on our previous studies of carbon solubility in nickel NPs \cite{diarra12,diarra12_2} to calculate the nickel rich side of the phase diagram of Ni-C NPs for systems up to 807 Ni atoms, for face centered cubic ($fcc$, Wulff-shaped) and $I_h$ NPs. We show that, beyond the well-known down shift of the melting points and liquidus lines, the two-phase solid / liquid domain of the bulk phase diagram is replaced by a solid core / liquid shell domain of the NP that extends to fairly low temperatures. We also evidence qualitative changes in the topology of the phase diagrams, related to the absence of a well-defined eutectic line.\\
  We developed an original approach, based on a carefully assessed atomic interaction model for Ni and C, together with computer simulation techniques that have already been discussed in \cite{diarra12,amara08,amara09}, so that only essential features are recalled here. To describe the Ni-C interactions keeping a quantum mechanical based framework, essential for safe parameterization and   transferability, we use a tight binding model that incorporates a moments description of local electronic densities of states at the minimal (4$^{th}$ moment) level, ensuring a linear scaling of the CPU time with system size. To calculate carbon incorporation in the NPs, with carbon chemical potential ($\mu_C$) as the control variable, we use Grand Canonical Monte Carlo (GCMC) \cite{frenkel} simulations. Since we  focus on carbon solubility in NPs, we  checked (see Supplemental Material \cite{SI}), that our model correctly reproduces it in bulk systems, with a maximum solubility around 5\% at 1865 K. Let us recall that the temperature scale has been reduced by a factor 0.85 to recover the experimental bulk melting temperature of Ni \cite{los10}.\\
 We studied the phase diagrams of $fcc$  NPs with 201, 405 and 807 Ni atoms in their equilibrium Wulff shape, and also considered icosahedral ($I_h$) structures with 55, 147 and 309 Ni atoms that present interesting differences. We used and complemented our previous calculations \cite{diarra12}, and the convergence of calculations on NPs with 405 and 807 atoms has been reassessed. Computational details, as well as all carbon incorporation isotherms used to calculate the phase diagrams are presented in Supp. Mat.  \cite{SI}. As observed previously, smaller NPs incorporate larger carbon fractions ($x_C$) at given $\mu_C$. The role of temperature ($T$) is similar and the solubility limit slightly increases with $T$. It is defined as the point in ($\mu_C$, $T$), and consequently ($x_C$, $T$) coordinates, where C atoms start segregating at the surface of the NP. A visual inspection of the NPs shows that increasing $\mu_C$ (and consequently $x_C$) induces a gradual melting of the NPs, that starts on the surface and propagates to the core.  Monte Carlo simulations do not yield diffusion coefficients, meaning that amorphous or liquid states cannot be rigorously discriminated. Since we wish to calculate the limits where carbon rich NPs transform into solid, core/shell or homogeneous liquid states, and locate the solubility limits to draw phase diagrams, we need to quantitatively define the molten and crystalline fractions of each NP.  This is done using the orientational order parameter first introduced by Steinhardt \emph{et al.} \cite{steinhardt83}, denoted $S_{i}$, that enables discriminating solid or liquid environments, for each atom i, see  \cite{SI}.
Averaging over all Ni atoms of the NP enables one to assign a global degree of crystallinity to the NP ($\bar{S}$). It is plotted in Fig. \ref{fig:fig1} for different temperatures and carbon compositions in the case of the Ni$_{309}$ $I_h$ NP.  We first note that the solid / liquid transition is gradual, with a linear dependence of $\bar{S}$ as a function of $\mu_C$. This appears to be a characteristic feature of finite size systems, since in an bulk system, liquid and solid phases should co-exist at the same chemical potential. The NP is considered crystalline for $\bar{S} \geq 0.85$, and disordered for  $\bar{S}\leq 0.35$. Between these values, solid core / molten shell structures prevail. As detailed in  \cite{SI}, plots similar to Fig. \ref{fig:fig1} are used to locate the transition points and size dependent ($x_C$, $T$) phase diagrams are readily obtained.\\
  We start by analyzing the phase diagram of the largest (807 Ni atoms) NP, displayed in Fig. \ref{fig:fig2}. At high $T$, calculations are easily converged and homogeneous solid (V) or liquid (I) NPs are identified, as well as a domain (II) where C segregates at the surface of a homogeneous liquid NP, forming chains that are highly mobile, sometimes partly detached from the NP surface.   As for a bulk system, the eutectic point (E) of the NP can be taken as the composition ($x_E$) at the lowest temperature ($T{_E}$) of liquid stability. Different from the bulk two phase domain, a solid core / liquid shell area (IV) continuously connects solidus and liquidus lines. Visual inspection, as well as the evolution of the order parameter as a function of $x_C$ and $T$ indicate that the liquid outer layer grows continuously at the expense of the solid with increasing carbon content. In the finite sized NP the two phases have to coexist, forming a rather well-defined interface. Within the limits of our calculations that are more difficult to converge at low $T$, and more difficult to analyze when disorder affects one or two Ni layers only, core / shell structures appear to be present below $T{_E}$. Indeed no sign of a well defined change at $T{_E}$ could be noticed. Thus, the solubility limit line is extended to lower temperatures. Because of the finite size of the system, interfaces are as important as the homogeneous parts of the system, and the Gibbs phase rule may not  apply. This is why a domain (III) is drawn, -separated from (II) and (IV) by dashed lines to indicate that the limits are rather fuzzy-, where carbon segregation takes place from an essentially solid NP, where disorder is limited to the outermost Ni layer and/or edges or vertices. More details are given in \cite{SI}.
  Let us note that the liquid phase remains stable down to about $x_C \sim$ 0.21 and  $T \sim 1020 K$, some $700 K$ below the bulk melting temperature ($1728 K$). The size reduction has a more dramatic effect on the carbon rich NP than on the pure one. Indeed, as compared to the bulk, the melting temperature of the pure Ni$_{807}$ NP is lowered by $\sim$10\%, while the eutectic temperature is decreased by $\sim$40\%. Contrary to the bulk case, though, no evidence for a well-defined isothermal equilibrium line, especially with a crystalline NP, can be found at $T_E$. On the contrary, an extension of the carbon solubility limit to temperatures below $T_E$ makes sense, since carbon segregation is observed  (see \cite{SI}) for $x_C$ $< 0.20$, from NPs where facets are still visible, while the outermost layer is disordered. The fairly low temperature ($\sim680 K$) required to observe C segregation from a faceted and essentially crystalline Ni NP of about 3 nm diameter makes it irrelevant for the selective growth of SWNTs, if we accept the existence of crystalline facets as a key factor to chiral selectivity \cite{yang14}.\\
  We now study the size dependence of the computed phase diagrams and compare them to the experimental bulk one \cite{portnoi10}, as presented in Fig. \ref{fig:fig3}. Initially $I_h$, with 55, 147 and 309 atoms, and $fcc$ Wulff shaped, with 201, 405 and 807 Ni atoms NPs are considered. In the presence of C, the outer layers become topologgically disordered, and the difference between $I_h$ and $fcc$ structures gradually washes out. The general trend is that, upon size reduction, the melting points and liquidus lines are shifted to lower temperatures. At constant temperature, the homogeneous liquid domain is larger for smaller NPs because the core / shell domain shrinks and carbon solubility limits of liquid NPs are shifted to larger C concentrations. Conversely, this means that, at a given temperature, one needs to reach larger carbon fractions to melt larger NPs. Although determining the solvus lines is difficult, a trend towards an enlargement of the solid state domain with increasing size is observed. Since C incorporation in the interstitial sites of the Ni lattice induces strain, the crystalline structure of smaller NPs is more easily destroyed by inserted C atoms, resulting in a vanishing solid state C solubility. A direct comparison of the solid state solubility limits of $I_h$ and $fcc$ NPs is difficult because sizes do not match. However, we do see that C incorporation is easier in Wulff shaped NPs, essentially because $fcc$ octahedral interstitial sites are larger than the tetrahedral ones present in the $I_h$ lattice, enabling C insertion with less strain, and also because (100) facets of the Wulff shape are more prone to adsorb and incorporate carbon.
  Beyond the detailed analysis of the structures, our results call for three general comments.\\
   A key feature of the Ni-C NPs phase diagrams is that NP melting is strongly favored by C incorporation, and proceeds from the surface to the core. This is best illustrated by the Ni-Ni and Ni-C pair distribution functions, whose evolution as a function of $\mu_C$ is presented in Fig. \ref{fig:fig4} at $850 K$ for Ni$_{309}$ and Ni$_{405}$ NPs. At low $\mu_C$ no or very few C atoms are incorporated and the crystalline structure of the NP is well preserved. With increasing $\mu_C$, C atoms are incorporated in the outer layers, inducing a gradual melting, characterized by blurred outer Ni layers. This results in a large downshift of the eutectic point, - by about 700 K for a Ni$_{807}$ $fcc$ NP - as compared to the bulk phase diagram. Such a behavior might be specific to the Ni-C system where carbon atoms are preferably incorporated in slightly too tight interstitial sites, but other transition metals or alloys might display the same feature.\\
  A second aspect, connected to the melting process, but possibly more general, is the large stability domain of solid core / liquid shell NPs that replaces the two-phase solid / liquid domain of the bulk phase diagram. While the coexistence of ordered crystalline and disordered (amorphous or liquid) parts of the NP is not that unexpected, the persistence of such core shell structures at temperatures below the eutectic point for small NPs below 3 nm indicates that a bulk like behavior, with an invariant eutectic equilibrium straight line joining the eutectic point and two solid phases is only recovered for larger NPs. Without being able to prove it by calculating a phase diagram for larger NPs, because of its heavy computing burden, we can easily envision the possibility that the extrapolation of the liquid solubility limit line below the eutectic point, separating domains III and IV in Fig. \ref{fig:fig2}, gradually tends to a horizontal line joining the eutectic point to a solid Ni-rich NP when its size grows bigger.\\
  Finally, turning to a practical application of Ni-C NPs to the synthesis of SWNTs, we see that, in the NP size range spanned here, below 3nm, the growing tubes are most probably in contact with a liquid or amorphous layer at temperatures relevant for CCVD. In principle, the tube cap nucleation takes place at or beyond the C saturation line, where the outer layer(s) of the NP is (are) disordered.  The carbon concentration within this disordered layer depends not only on temperature, but also on its chemical potential ($\mu_C$), determined by the thermochemistry of the decomposition reaction of the carbon feedstock.  We know that the contact angle between the Ni-C NP and a growing nanotube depends on the fraction of C dissolved in the NP \cite{diarra12_2}, leading to different growth modes \cite{fiawoo12}. For low carbon fractions, the Ni NPs tend to wet the C $sp^2$ walls, so that the growing tube and the NP have essentially the same diameter (tangential growth), while larger carbon fractions lead to dewetting, so that the tube has a smaller diameter than the catalyst NP (perpendicular growth). In addition, related to the possibility of a selective growth of SWNTs from solid NPs, our calculations on Ni indicate that the eutectic temperature is strongly lowered as compared to the bulk one. For the Ni$_{807}$ NP, a homogeneous liquid phase with $\sim$21\% C, is stable down to $\sim$1020 K, that is $\sim$60\% of the bulk melting temperature. If we extrapolate this to another alloy of interest such as the W$_6$Co$_7$ compound \cite{yang14},  that melts around $2100 K$ in the bulk, we see that the question of its physical state (solid, core / shell or liquid) under growth conditions, around $1200 K$, remains open. This, of course, depends on the size of the catalyst NP. We thus understand how the detailed understanding and predictive evaluation of the state of the catalyst NP, translated in the form of a size dependent phase diagram, can contribute to a better controlled and possibly selective growth of SWNTs.

\begin{acknowledgements}
The research leading to these results has received funding from the European Union Seventh Framework Programme (FP7/2007-2013) under grant agreement n° 604472. The authors also acknowledge support from the French Research Funding Agency (ANR-13-BS10-0015-01)
\end{acknowledgements}

\end{document}


\title{Supplemental Material for "Size dependent phase diagrams of Nickel-Carbon nanoparticles"}

\author{Y. Magnin}
\affiliation{Centre Interdisciplinaire de Nanoscience de Marseille, Aix-Marseille University and CNRS, Campus de Luminy, Case 913, F-13288, Marseille, France.}

\author{A. Zappelli}
\affiliation{Centre Interdisciplinaire de Nanoscience de Marseille, Aix-Marseille University and CNRS, Campus de Luminy, Case 913, F-13288, Marseille, France.}

\author{H. Amara}
\affiliation{Laboratoire d'Etude des Microstructures, ONERA and CNRS, BP 72, F-92322, Ch\^atillon, France.}

\author{F. Ducastelle}
\affiliation{Laboratoire d'Etude des Microstructures, ONERA and CNRS, BP 72, F-92322, Ch\^atillon, France.}

\author{C. Bichara}
\affiliation{Centre Interdisciplinaire de Nanoscience de Marseille, Aix-Marseille University and CNRS, Campus de Luminy, Case 913, F-13288, Marseille, France.}

\maketitle

\begin{center}
\begin{figure}[htb!]
\includegraphics[width=14cm]{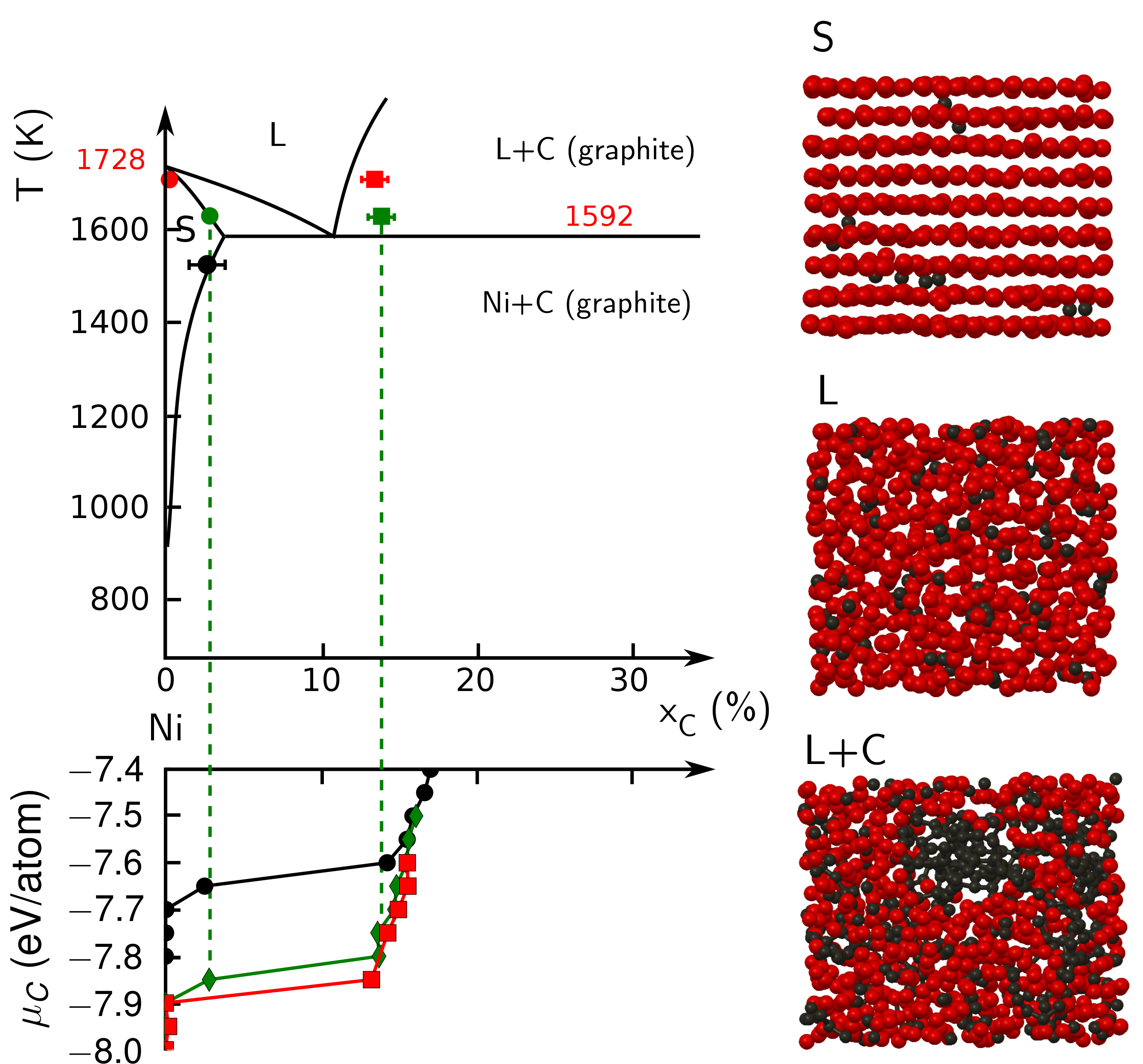}
  \caption{(Color online) Carbon incorporation isotherms in a 576 Ni bulk system (lower figure) and the corresponding phase boundary points of the phase diagrams, compared to the experimental one \cite{portnoi10} (full black lines, upper figure). The red, green and black circles on the phase diagram correspond to the carbon solubility limit in the solid at 1700, 1615 and 1530 K respectively. The red and green squares indicate the C concentration in the liquid, at the next carbon chemical potential step, 0.05 eV/atom higher. Calculations at 1530 K and -7.60 eV/atom did not show a convergence of the C concentration, but the simulation box was clearly disordered: this is the reason why it is not shown on the phase diagram. Three typical situations (solid with C dissolved, homogenous liquid and carbon segregation from the liquid) are depicted on the right.}
  \label{fig:fig1}
\end{figure}
\end{center}

\textbf{Check of the solubility in bulk Nickel}\\
Since we will focus on carbon solubility in nanoparticles, we checked that our model correctly reproduces it in bulk systems. Because carbon incorporation in interstitial sites causes an expansion of the system \cite{Moors09}, we have to use so-called "osmotic ensemble" Monte Carlo simulations \cite{Coudert08}. In our case, the total number of Ni atoms, carbon chemical potential, (zero-) external pressure and temperature are kept constant. Because of the non-trivial atomic interaction model used \cite{amara09}, these calculations are extremely difficult to converge and large hysteresis are observed, starting from either pure solid Ni or C-rich liquid alloy. To get a reasonable order of magnitude of C solubility, very long simulation runs had to be performed, starting from a mixed configuration with a C-rich liquid nucleus inside an essentially solid Ni-C dilute alloy. Only three temperatures of the Ni-C phase diagram could be calculated. We then see in Supp. Mat. Figure \ref{fig:fig1} that the calculated maximum carbon solubility in solid bulk Ni, is around 5\%, in quite good agreement with the experimental one.\\

\begin{center}
\begin{figure}[htb!]
\includegraphics[width=15cm]{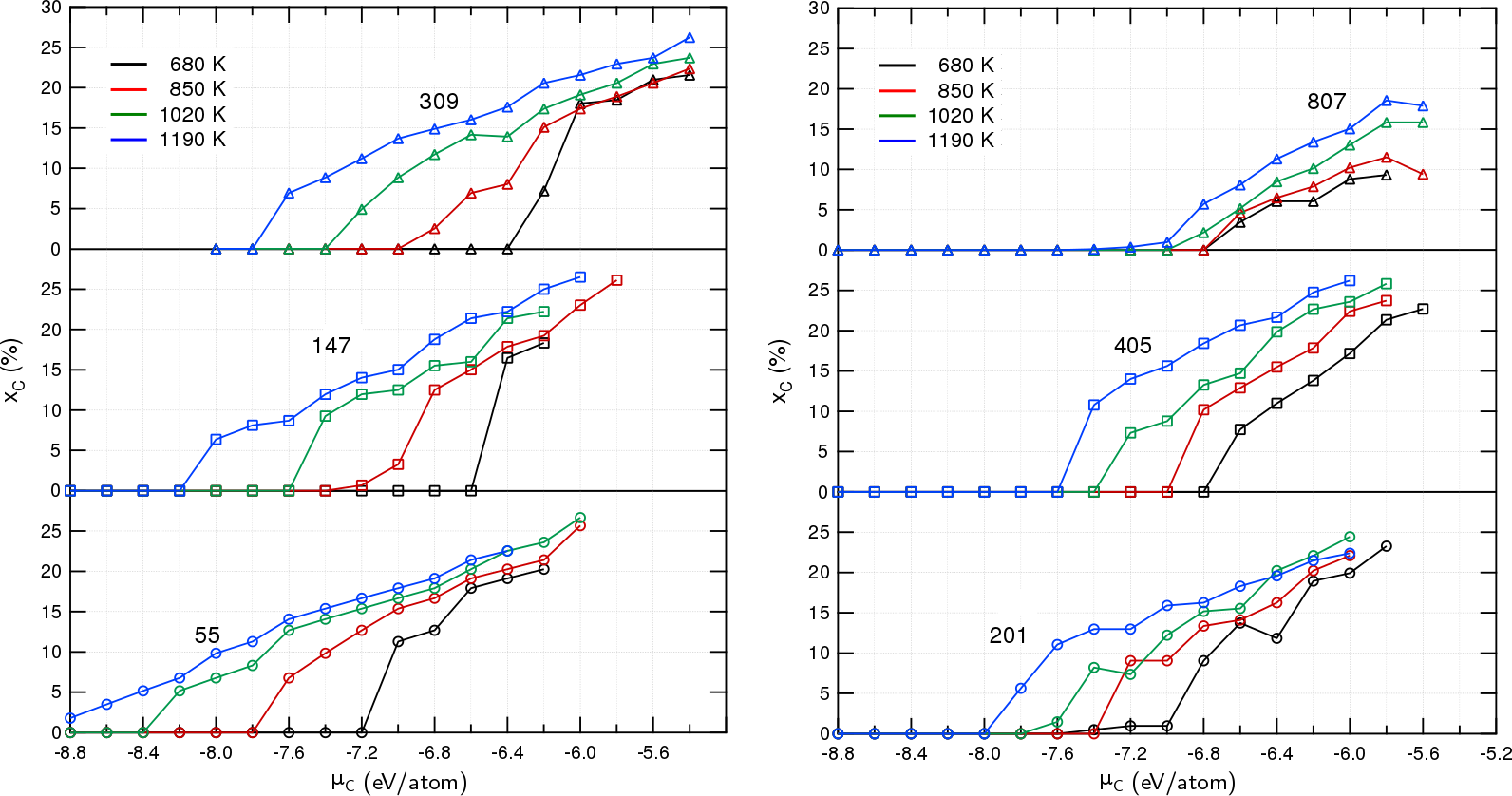}
  \caption{(Color online) Carbon sorption isotherms on icosahedral (left), with 55, 147 and 309 Ni atoms, and face centered cubic, Wulff shaped (right), with 201, 405, 807 Ni atoms nanoparticles. These curves present the carbon fraction (x$_C$) inside the nanoparticle, as a function of temperature and C chemical potential ($\mu_C$). Low values of $\mu_C$, lead to a small fraction of C dissolved. Increasing $\mu_C$, the C fraction (x$_C$) continuously grows, while the NP gradually melts. At constant $\mu_C$, the fraction of dissolved C is larger at higher temperatures.}
  \label{fig:fig2}
\end{figure}
\end{center}

\textbf{Monte Carlo calculation of the carbon sorption isotherms on Nickel NPs}\\
The Grand Canonical Monte Carlo algorithm consists here in a series of macro-steps. Each macro-step randomly alternates displacement moves for Ni or C atoms, attempts to incorporate carbon in a previously defined active zone and attempts to remove existing carbon atoms. In order to mimic the CCVD process, the active zone for inserting or extracting carbon atoms is defined as a region of space at less than 0.3 nm above and 0.3 nm below the surface of the Ni cluster. We typically performed up to 5000 macro-steps, but sometimes had to double this number. Within each macro-step, we systematically achieved four times the number of atom attempted displacement steps. To incorporate (resp. remove) carbon atoms in the structure, 1000 attempts were made and the corresponding routine was exited as soon as the first successful incorporation (resp. deletion) occurred. We usually started from pristine crystalline Ni NPs, but checked that the result was stable against a change of the initial configuration. We also checked that the amount of C adsorbed did not significantly depend on the choice of the thickness of the incorporation active zone, at least for the smallest NP sizes. Because of the large surface/bulk ratio, hysteresis problems are less critical than in the bulk. However, calculations turn out to be extremely long for the biggest NPs (up to 3 months for 807 Ni atoms and large fractions of C incorporated). This might lead to systematic underestimation of the C fractions. For smaller NPs fluctuations are more important, leading to some statistical noise, especially in the case of core / shell NPs.\\
Once the equilibrium is reached, the number of C incorporated in the NP fluctuates around an average value. We record the number of C atoms adsorbed inside the cluster at given $\mu_C$ and T. Quite obviously, once the NP is saturated with carbon, for high values of $\mu_C$, C atoms are also stabilized on the surface or outside the NP. We do not consider them in the present study. In Supp. Mat. Figure \ref{fig:fig2}, we present the C sorption isotherms at 4 temperatures: 680, 850, 1020 and 1190 K. Some of the already published data \cite{diarra12} are presented here for the sake of completeness. New calculations include the icosahedral NP with 309 Ni atoms. The convergence of the runs with 405 and 807 was also improved.\\

\textbf{Analysis of the solid / liquid state of the Nanoparticles}\\
The atomic structure of the nanoparticles is then investigated using the local order parameter (S$_i$) proposed by Steinhardt et \textit{al.} \cite{steinhardt83}, and improved by Jungblut et \textit{al.} \cite{jungblut13}, equation (\ref{eq:op1}), to define the ordered or disordered (i.e.: amorphous or liquid) parts of the nanoparticles. This parameter reflects the symmetry of the local environment of each atom, making use of spherical harmonics. It writes:
\begin{eqnarray}
\label{eq:op1}
S_{i}&=&\frac{1}{N_b} \displaystyle \sum^{N_b}_{j=1}\frac{\displaystyle \sum^{6}_{m=-6}q_{6m}(i)q^*_{6m}(j)}{\left(\displaystyle \sum^{6}_{m=-6}|q_{6m}(i)|^2\right)^{1/2} \left(\displaystyle \sum^{6}_{m=-6}|q_{6m}(j)|^2\right)^{1/2}}
\end{eqnarray}
where N$_b$ is the number of neighbors j of atom i,
\begin{eqnarray}
\label{eq:op2}
q_{6m}(i)&=&\frac{1}{N(i)}\sum^{N(i)}_{j=1}Y_{6m}\left(\theta(\textbf{r}_{ij}),\phi(\textbf{r}_{ij})\right)
\end{eqnarray}
and Y$_{lm}$ is the spherical harmonic of degree l and order m.\\

\noindent Averaging over the Ni atoms of the NP yields a global order parameter $\bar{S}$, used in the following to establish the phase diagram. $\bar{S}$ is normalized in such a way that a perfectly crystallized structure has $\bar{S}$=1, while a fully disordered structure corresponds to $\bar{S}$=0. These are limiting values and we practically consider structures with $\bar{S}\geq$ 0.85 as solid and those with $\bar{S}\leq$ 0.35 as liquid.\\

\begin{center}
\begin{figure}[htb!]
\includegraphics[width=12cm]{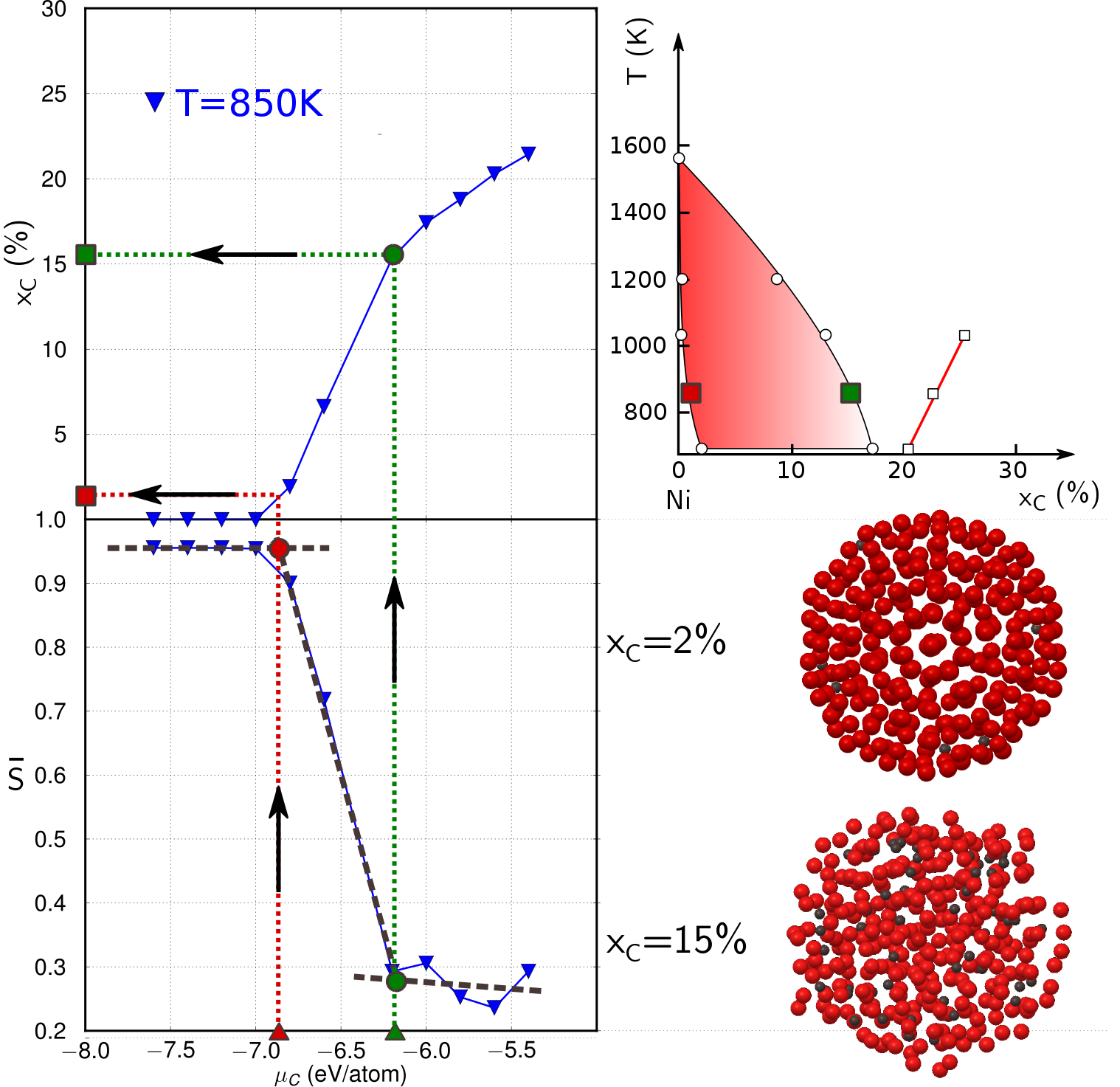}
  \caption{(Color online) Determination of phase boundaries of an icosahedral particle with 309 Ni atoms. The liquid ($\bar{S}<$ 0.35) and solid ($\bar{S}>$  0.90) boundaries are determined in the ($\bar{S}$, $\mu_C$) plane (lower panel) and the corresponding concentrations are then obtained from the C incorporation isotherm (upper panel) and located in the phase diagram.}
  \label{fig:fig3}
\end{figure}
\end{center}

\textbf{Phase diagram calculation}\\
The first step is to plot the global order parameter ($\bar{S}$) as a function of $\mu_C$, at different temperatures. A typical result is presented in Supp. Mat. Figure \ref{fig:fig3} that presents the transition from a crystalline NP, with $\bar{S}\sim$ 0.95 for $\mu_C\leq$ -7.0 eV/atom to a disordered one, with $\bar{S}\leq$ 0.30 for $\mu_C\geq$ -6.2 eV/atom. In the intermediate region (-7.0 $\leq\mu_C\leq$ -6.2 eV/atom), we note that $\bar{S}$ depends linearly on $\mu_C$. This is a characteristic feature of the "phase transition" in a nano-sized system that continuously evolves from solid to liquid by forming liquid shell / solid core nanoparticles. In a large bulk system, the solid / liquid transition would take place at a fixed $\mu_C$ value. The intersections of the three pieces of straight lines yield the carbon chemical potentials at the transition points that are readily translated in carbon concentration using the carbon sorption isotherm. The corresponding solid and liquid points are then reported on the phase diagram. This procedure is repeated at different temperatures and for different nanoparticle sizes to obtain the solvus and liquidus lines of the phase diagrams.\\

\begin{center}
\begin{figure}[htb!]
\includegraphics[width=18cm]{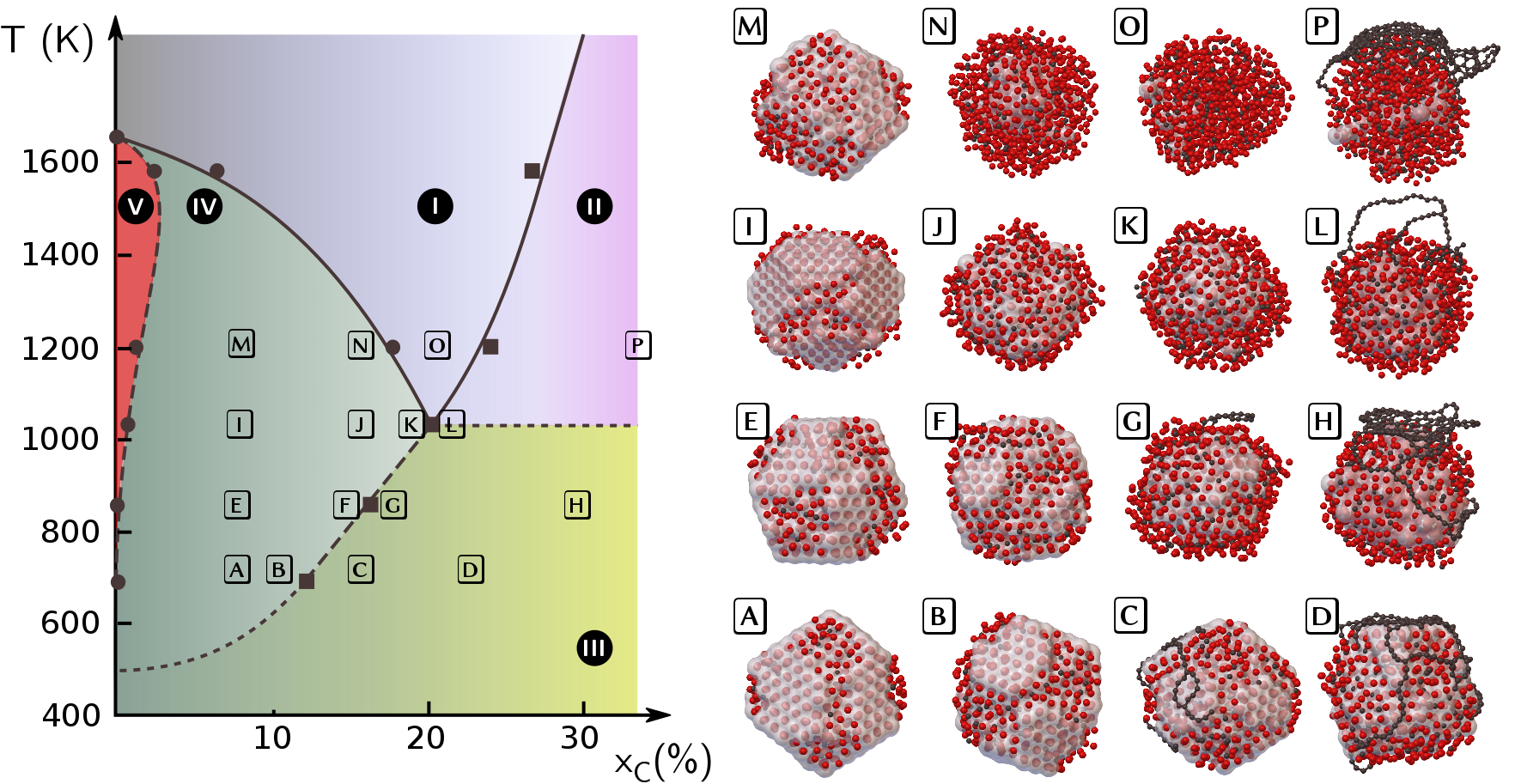}
  \caption{(Color online) Left panel: Ni-C phase diagram for a nanoparticle with 807 Ni atoms.\\
The red domain (V), limited by a dashed solvus line, corresponds to a homogenous solid solution. The blue domain (I), limited by liquidus lines (full), corresponds to homogenous liquid NPs. The left green region (IV), between V and I, corresponds to solid core / liquid shell nanoparticles. On the right of the phase diagram, a segregation of carbon at the surface of the NP is observed, on liquid (purple, II) or core / shell (yellow green, III) nanoparticles. The dashed line separating domains III and IV at temperatures below 700 K is hypothetical, while the dashed line between 700 and 1000 K, is an extrapolation of the solubility limit into the core / shell domain, supported by an analysis of the atomic configurations. 
Right panel: images of the NPs located in the left panel phase diagram by the position of their identification letter. 
The grey surface is a contour of the solid core of the NP, if it exists. Red and black balls respectively correspond to Ni atoms with a small $\bar{S}>$ parameter ("liquid") and their surrounding C atoms.}
  \label{fig:fig4}
\end{figure}
\end{center}

\textbf{Detailed presentation of the Phase diagram for 807 Nickel Nanoparticle}\\
In order to better understand how the phase diagrams were built, we provide more visual information on the structure of the nanoparticles with 807 Ni atoms and different carbon fractions. Results are shown in Supp. Mat. Figure \ref{fig:fig4}, that presents 16 snapshots of NPs at different locations in the (x$_C$, T) phase diagram. In particular, along the M, I, E and A configurations that contain $\sim$ 8\% C at temperatures between 1200 and 700 K, we see a continuous evolution in the core shell structures, without any significant change at the eutectic temperature. This supports the idea that the liquid C-saturation line can be extrapolated below the eutectic temperature and, consequently, that no 3 phase equilibrium line exists in small NPs.

